\newcommand{\PaperNumber}{154, 6 pages body}

\documentclass[sigconf]{acmart}

\copyrightyear{2024}
\acmYear{2024}
\setcopyright{acmlicensed}
\acmConference[IMC '24] {Proceedings of the 2024 ACM Internet Measurement Conference}{November 4--6, 2024}{Madrid, Spain.}
\acmBooktitle{Proceedings of the 2024 ACM Internet Measurement Conference (IMC '24), November 4--6, 2024, Madrid, Spain}
\acmISBN{979-8-4007-0592-2/24/11}
\acmDOI{10.1145/3646547.3689011}

\usepackage{tikz}
\usepackage{multirow}
\usepackage{amsmath,amsfonts}

\usepackage{paralist}
\usepackage{url}

\usepackage{threeparttable}
\usepackage{booktabs}
\usepackage{makecell}
\usepackage{wasysym}
\usepackage{enumitem}
\usepackage{colortbl}
\usepackage{algorithm}
\usepackage[noend]{algpseudocode}
\usepackage{graphicx}
\usepackage{subcaption}
\usepackage{float}
\usepackage{xcolor}
\usepackage{xspace}

\def\BibTeX{{\rm B\kern-.05em{\sc i\kern-.025em b}\kern-.08em
    T\kern-.1667em\lower.7ex\hbox{E}\kern-.125emX}}

\newcommand{\ignore}[1]{}

\newcommand{\system}{\textsc{NetDPSyn}\xspace}

\begin{document}

\title{\system: Synthesizing Network Traces under Differential Privacy}


\author{Danyu Sun}
\orcid{0009-0001-0066-3459}
\affiliation{%
\institution{University of California, Irvine}
\city{Irvine}
\state{CA}
\country{USA}
}
\email{danyus2@uci.edu}

\author{Joann Qiongna Chen}
\orcid{0009-0009-0787-058X}
\affiliation{%
\institution{San Diego State University}
\city{San Diego}
\state{CA}
\country{USA}
}
\email{jchen27@sdsu.edu}

\author{Chen Gong}
\orcid{0000-0001-6178-4118}
\affiliation{%
\institution{University of Virginia}
\city{Charlottesville}
\state{VA}
\country{USA}
}
\email{fzv6en@virginia.edu}

\author{Tianhao Wang}
\orcid{0000-0002-9017-7947}
\affiliation{%
\institution{University of Virginia}
\city{Charlottesville}
\state{VA}
\country{USA}
}
\email{tianhao@virginia.edu}

\author{Zhou Li}
\orcid{0000-0002-9401-1012}
\affiliation{%
\institution{University of California, Irvine}
\city{Irvine}
\state{CA}
\country{USA}
}
\email{zhou.li@uci.edu}

\begin{abstract}
As the utilization of network traces for the network measurement research becomes increasingly prevalent, concerns regarding privacy leakage from network traces have garnered the public's attention. To safeguard network traces, researchers have proposed the \textit{trace synthesis} that retains the essential properties of the raw data. 
However, previous works also show that synthesis traces with generative models are vulnerable under linkage attacks. 

This paper introduces \system, the first system to synthesize high-fidelity network traces under privacy guarantees. \system is built with the Differential Privacy (DP) framework as its core, which is significantly different from prior works that apply DP when training the generative model.
The experiments conducted on three flow and two packet datasets indicate that \system achieves much better data utility in downstream tasks like anomaly detection. \system is also 2.5 times faster than the other methods on average in data synthesis.
\end{abstract}



\maketitle
\pagestyle{plain}

 \section{Introduction}
\label{sec:intro}

Network traces are critical for guiding and designing many applications such as network telemetry and network-based anomaly detection.
However, prominent privacy risks of sharing such data have to be considered, as the shared information might identify a participant or even leak sensitive attributes. 
The conventional approaches to contain information leakage from the published network traces mainly follow data redaction like IP anonymization~\cite{xu2002prefix}. However, such methods are vulnerable under inference attacks~\cite{dwork2017exposed} and data reconstruction attacks~\cite{garfinkel2019understanding, cohen2018linear}. 
An alternative approach is to \textit{synthesize} traces that capture the properties from the raw traces, gaining strong momentum as it addressed privacy regulations.
Recently, Generative Adversarial Network (GAN) has been leveraged to synthesize network traces~\cite{yin2022practical, lin2020using, fan2021dpnet, sivaroopan2023synig, ring2019flow, cheng2019pac, wang2020packetcgan, han2019packet}. 
However, there is no \textit{guarantee} that the users' privacy behind the traces is well guarded, and Stadler et al. showed that the data synthesized under the generative models suffer from linkage attacks~\cite{stadler2022synthetic}.

Differential Privacy (DP)~\cite{dp}, which incorporates calibrated noise to ``hide'' the existence of an individual's data, shows promise in providing the essential privacy guarantees against the aforementioned attacks. 
There has been early effort to add DP noises to the responses according to the \textit{statistical queries}~\cite{mcsherry2010differentially}, but the traces were not released.
Some recent works enhance the GAN-based trace synthesis with DP~\cite{yin2022practical, fan2021dpnet}, mainly through
DP-SGD~\cite{abadi2016deep}. 
However, the data utility is significantly worsened even under a very relaxed privacy budget (see Section~\ref{subsec:motivation}). 
{\it ``Research on privacy-preserving network data sharing using rigorous approaches such as differential privacy is needed''}~\cite{imana2021institutional}, but researchers also recognized that  {\it``a critical gap remains identifying how these privacy-preserving technologies can be applied to networking problems''}~\cite{claffy2021workshop}.

In this work, we pursue a different direction from prior works~\cite{yin2022practical, fan2021dpnet}: instead of synthesizing network records with generative models that is trained under DP, we try to capture the underlying distributions of the original data and synthesize network records from them \textit{after} they are protected by DP. 
This design choice is driven by our insight that the underlying distributions are more critical to many downstream applications like anomaly detection, and by directly controlling these distributions under DP, we avoid the need for excessive noise addition when training the generative model. 
Specifically, we develop a new system \system that extends a marginal-based synthesizer, PrivSyn~\cite{zhang2021privsyn}, for the network settings.

We conducted a comprehensive study to evaluate the effectiveness of \system, by comparing with 3 baseline methods on 5 datasets. Our initial result shows the data synthesized by \system can achieve similar fidelity even as the raw data for tasks like flow classification. \system is also much more efficient than the other baselines. Our code is available at \url{https://github.com/DanyuSun/NetDPSyn}.

\section{Background}
\label{sec:background}
\subsection{Network Dataset and Privacy}
\label{subsec:network}

Like previous works~\cite{yin2022practical, lin2020using}, we consider \textit{header} fields of network packets or flows as the target for data synthesis. Releasing the header fields is often sufficient to support many research scenarios like network-based anomaly detection~\cite{ids-dataset}, and addresses some privacy concerns (e.g., a payload might directly contain personal data). We use 5 public datasets that either contain packets or flows, as elaborated in Section~\ref{subsec:setup}. 
Below we describe their common fields.

\begin{itemize} [noitemsep,topsep=0pt]
    \item \textbf{Packet header.} It contains information about individual packets at layer 3 (IP layer) and layer 4 (transport layer). Specifically, it contains the source and destination IP addresses (\texttt{srcip} and \texttt{dstip}), source and destination ports (\texttt{srcport} and \texttt{dstport}), protocol type up to layer 4 (\texttt{proto}, e.g., TCP, UDP, and ICMP), timestamp of capture (\texttt{ts}), packet length (\texttt{pkt\_len}), and other fields like checksum (\texttt{chksum}) and label given by the data collector (\texttt{label}).
    \item \textbf{Flow header.} A network flow aggregates the packets under IP 5-tuple $\langle$\texttt{srcip}, \texttt{dstip}, \texttt{srcport}, \texttt{dstport}, \texttt{proto}$\rangle$~\cite{bagnulo2011stateful}. The timestamp of the first packet (\texttt{ts}), the duration of flow (\texttt{td}), the number of packets (\texttt{pkt}), the number of bytes (\texttt{byt}), and flow label (\texttt{label}) are also included per flow.
\end{itemize}


Releasing the header without the payload still raises privacy concerns, and the main solutions include data anonymization and data synthesis. Data anonymization is often performed on IP addresses, e.g., with CryptoPan prefix-preserving 
anonymization ~\cite{xu2002prefix}. Yet, a recent study showed such IP anonymization is still vulnerable if the institution associated with a prefix has sensitive Internet activities (e.g., sending email to a controversial organization)~\cite{imana2021institutional}. Data synthesis has finer-grained control over the privacy-utility tradeoff~\cite{yin2022practical}, 
and we aim to provide \textit{provable} privacy protection for the synthesized network traces while maintaining their data utility.

\subsection{Differential Privacy}
\label{subsec:dp}

Differential privacy (DP)~\cite{dp} guarantees that the result of any computation on a dataset, such as a database query, remains essentially unchanged whether or not any single individual's data is included or excluded.  

\subsubsection*{Definition 1} (\textit{Differential Privacy~\cite{dp}}) A randomized mechanism $\mathcal{A}$ satisfies ($\varepsilon, \delta$)-differential privacy, if and only if, for any two neighboring datasets $D$ and $D'$, it holds that,
\begin{equation}\label{eq:dp}
    \Pr[\mathcal{A}(D) \in \mathcal{O}] \leq e^\varepsilon \Pr[\mathcal{A}(D') \in \mathcal{O}] + \delta,
\end{equation}
where $\mathcal{O}$ represents the set of all conceivable outputs of the algorithm $\mathcal{A}$. The privacy budget $\varepsilon$ and $\delta$ are both non-negative parameters that indicate the privacy loss in the data. 
A lower value of $\varepsilon$ signifies enhanced privacy and a smaller $\delta$ corresponds to a decreased probability that the privacy protection assured by $\varepsilon$ will be broken.
The granularity of DP is dependent on how the neighboring dataset is defined. Typically, two datasets ${D, D'}$ differing in only one record are considered neighboring, which can be regarded as \textit{record-level} DP. 
For instance, this study defines the `record' as a single log entry of a network packet or flow. 

\subsection{Related Works}
\label{subsec:related}

\noindent \textbf{Synthesis of network traces.}
Based on our literature review, recent works prefer to synthesize network traces with generative models.
Among them, most of the works applied Generative Adversarial Networks (GANs)~\cite{yin2022practical, lin2020using, fan2021dpnet, sivaroopan2023synig, ring2019flow, cheng2019pac, wang2020packetcgan, han2019packet}, which trains \textit{generator(s)} to map a noise vector to a sample and \textit{discriminator(s)} to classify the sample as fake or real. 
NetShare is a prominent example~\cite{yin2022practical}, which has achieved high data fidelity on the header traces with a time-series generator.

Recently, 
a few works applied \textit{diffusion} models to synthesize network traces~\cite{jiang2023netdiffusion, sivaroopan2023netdiffus}.
Take NetDiffus as an example~\cite{sivaroopan2023netdiffus}. It converts network traces into a sequence of 2-D images, iteratively adds noises in the forward pass, and trains a denoiser to recover the original network traces.

\vspace{2pt} \noindent \textbf{DP for dataset synthesis.} 
DP has been extensively leveraged to construct synthetic data that can be shared under privacy guarantees~\cite{hu2023sok}. Different types have been tested, including image~\cite{dp-diffusion,dockhorn2023differentially,li2023meticulously,liew2022pearl}, tabular~\cite{zhang2021privsyn,privlava,privmrf,aim,yoon2018pategan,PrivBayes}, graph~\cite{privgraph}, time-series~\cite{dpserisdataset01}, trajectory~\cite{wang2023privtrace,du2023ldptrace}, text data~\cite{yue-etal-2023-synthetic,tang2023privacypreserving} and more.

\system treats network traces as tabular data and follows marginal-based synthesis. 
Another direction is copula-based synthesis~\cite{gambs2021growing}, which uses a copula function (e.g., Gaussian copula) to model the joint distribution and synthesize traces~\cite{gambs2021growing, asghar2020differentially, chanyaswad2019ron, li2014differentially}. We did preliminary experiments with Gaussian copula, but the result was unsatisfactory. Using a different copula function or adapting the Gaussian copula for the network datasets might be needed, and we leave it as a future work.

\subsection{Threat Model}
\label{subsec:threat}

We follow the threat modeling of Houssiau et al.~\cite{houssiau2022tapastoolboxadversarialprivacy}: given a dataset synthesized from the original dataset, the attacker is motivated to conduct 3 types of attacks, including membership inference attack (MIA),  attribute inference attack (AIA) and data reconstruction attack. MIA aims to determine whether a data point is included by or excluded from a synthetic dataset. AIA aims to infer sensitive attributes from the other released attributes. Data reconstruction aims to recover the entire original dataset. DP is supposed to deter all these inference attacks by modeling the worst-case scenario (i.e., strongest attacker)~\cite{de2024synthetic}.

In Appendix \ref{app:mia}, we provide a preliminary evaluation of the effectiveness of the basic MIA method~\cite{yeom2018privacy} on the synthesized network traces, suggesting DP is an effective defense.

\section{\system}
\label{sec:design}

\subsection{Motivation and Workflow}
\label{subsec:motivation}

We revisit the generative-model-based approaches for trace synthesis and argue that they cannot rigorously guard users' privacy.
In fact, Stadler et al. found the generative models like CTGAN~\cite{xu2019modeling} suffer from \textit{linkage attacks}, which allows an attacker to infer the presence of a record in the original dataset with high confidence~\cite{stadler2022synthetic}. 
Though DP has been considered to harden the generative models~\cite{yin2022practical, fan2021dpnet}, we found 
they all choose DP-SGD~\cite{abadi2016deep}, which clips each gradient and adds Gaussian noise to make the generative models fulfill $(\epsilon, \delta)$-DP. 
However, the data utility can be significantly worse. For example, Table 5 of ~\cite{yin2022practical} shows the Earth Mover’s Distance (EMD) \textit{is increased from 0.10 (without DP-SGD) to 0.35 (with DP-SGD), even under a very large $\varepsilon = 24.24$}. In fact, DP-SGD is known to add excessive noises due to that its privacy guarantee is proved on \textit{each SGD step}~\cite{feldman2018privacy}.

\begin{figure}[t]
  \centering
  \includegraphics[width=\linewidth]{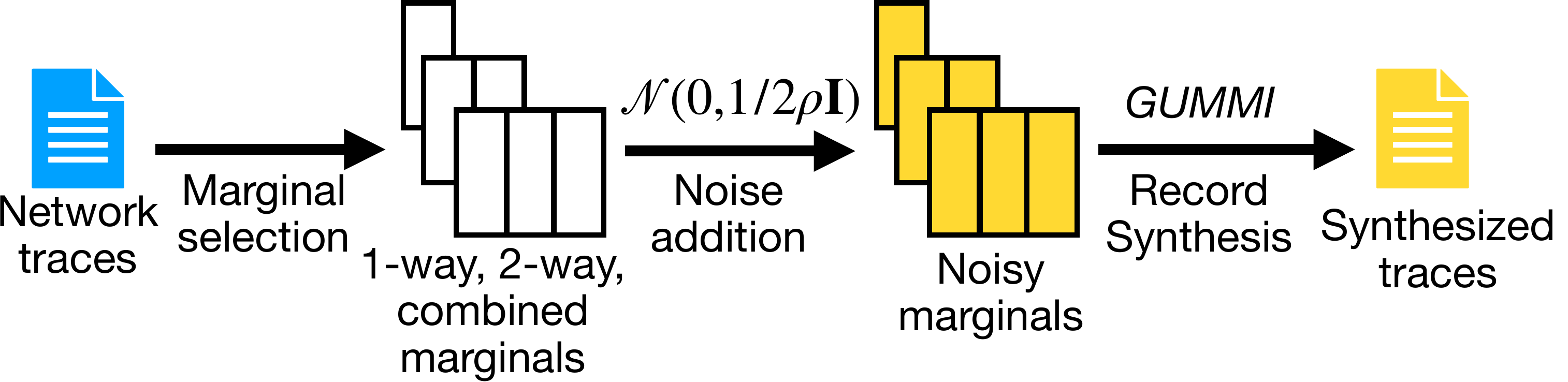}
  \caption{High-level workflow of \system.}
  \label{fig:workflow}
  \vspace{-10pt}
\end{figure}

Instead of synthesizing network records with generative models and modifying them to fulfill DP, we choose to capture the underlying distributions of the original data and synthesize network records from them after DP protects them. By doing so, we can directly control the privacy-utility tradeoff on the data distributions and avoid adding excessive noises.
Figure~\ref{fig:workflow} shows the general workflow of our approach. 

To generate DP-protected correlations, we extend a prior work PrivSyn~\cite{zhang2021privsyn} which handles high-dimensional datasets by  automatically selecting and constructing noisy marginal tables that capture the data distributions. It turns out to be more effective and efficient than the approaches based on probabilistic graphical model (e.g., Bayesian network~\cite{PrivBayes} and Markov Random Fields~\cite{mckenna2021relaxed, PGM,aim}) as shown in our evaluation. We incorporate the \textit{network domain knowledge} to adapt PrivSyn for network traces, by 1) developing a new binning technique to pre-process continuous attribute values; 2) augmenting the network traces with temporal features; 3)  applying various protocol constraints for data consistency; 4) bootstrapping Gradually Update Method (GUM) of PrivSyn with an initialized dataset to scale up the synthesis process.


\system consists of 3 main steps. First, it pre-processes the network traces with data binning and feature addition. Then, it selects 2-way marginals under an optimization procedure, publishes them by adding Gaussian noises, and post-processes the marginal tables to apply protocol-related restrictions for consistency. Finally, it synthesizes the network traces from the marginal tables. 
The workflow of \system with pseudo-code is shown in Appendix~\ref{app:workflow}.

\subsection{Pre-processing}
\label{subsec:pre-processing} 

Fields with large domain sizes could lead to excessive noise under DP mechanisms like Gaussian mechanisms, and PrivSyn combines \textit{all} low-count values into a new value, but this approach would remove too many low-count values or merge values that are not relevant.  We found this approach is particularly problematic for network traces, and 
we propose a finer-grained method to bin different network fields disparately based on their distribution and types.

As the first step, we use a \textit{type-dependent} binning method on each attribute, and consider 5 field types. 1) IP  (\texttt{srcip} and \texttt{dstip}): we bin the low-count IP addresses by the \texttt{/30} prefix. 2) Port  (\texttt{srcport} and \texttt{dstport}): we keep a list of common ports under 1024 away from the binning process, and bin the higher port numbers by 10. 3) Categorical attributes (e.g., \texttt{proto} and \texttt{label}) with small domain size are not binned. 4) Integer and floating-point attributes (\texttt{pkt}, \texttt{byt}  and \texttt{td}): they are binned under log transformation, i.e., $log(1+x)$, which results in much smaller number of bins than linear binning. 5) Timestamp (\texttt{ts}): we handle them with a new method detailed in the end of this subsection.

After the type-dependent binning, we perform another round of frequency-dependent binning, to further aggregate the bins with small frequencies. As frequency is related to the actual raw data, to control the information leakage introduced in this step, we apply DP before making the bin selection. In particular, for each attribute, we publish noisy 1-way marginals by applying Gaussian Mechanism (see ``Adding Noise to Marginals'' in Section~\ref{subsec:marginals}) on the marginal cell count (e.g., $M_{\text{d}}(v)$ in Table~\ref{tbl:marginals} of Appendix~\ref{app:marginals}), with a portion of privacy budget.

\vspace{2pt} \noindent \textbf{Capturing temporal pattern.}
 
Inspired by~\cite{lin2020using}, we create another field \texttt{tsdiff} (the differences between timestamps) from the \texttt{ts} field, which will be treated the same as other fields under data binning,  marginal-table selection, etc. 
Instead of directly computing \texttt{tsdiff} on two adjacent time-ordered records, we group the records by an identifier, e.g., the IP 5-tuple, and compute the \texttt{tsdiff} in each group. We focus on the group-wise \texttt{tsdiff} as the activities of different groups  are less likely to correlate. 
We use \texttt{tsdiff} to represent the temporal patterns because it reflects packet-arrival intervals, which are widely used in  downstream applications, like autocorrelation~\cite{box2015time}.

\subsection{Generating Noisy Marginal Tables}
\label{subsec:marginals}

After the pre-processed 1-way marginal tables are generated, we can construct the 2-way marginal tables from them and publish their noisy versions using Gaussian Mechanism. The 2-way marginals capture the field correlations that are essential for high-fidelity record synthesis.
However, publishing \textit{all} 2-way marginal tables will introduce a large amount of DP noises, so we follow PrivSyn's  \textsc{DenseMarg} algorithm to select useful 2-way marginals under low privacy budget. Though data owners can manually select marginals to be released, we argue that it is very difficult to select the right subset for the best privacy-utility tradeoff, so we take a data-driven approach.
Examples of 1-way and 2-way marginal tables and their noisy versions  are shown in Appendix~\ref{app:marginals}.

\vspace{2pt} \noindent \textbf{Marginal selection.} \textsc{DenseMarg} formalizes the marginal selection problem as an optimization problem that balances dependency error (error caused by missing a marginal) and noise error (error caused by adding noises to a selected marginal) as below:
\setlength{\abovedisplayskip}{1pt}
\setlength{\belowdisplayskip}{1pt}
\begin{equation}\label{eq:selection}
\text{minimize} \sum_{i=1}^{m} [\psi_i x_i + \phi_i(1 - x_i)]
\text{ s.t. } x_i \in \{0,1\}  
\end{equation}
where $m$ is the number of all marginals ($m=d(d-1)/2$ for $d$ fields), $\psi_i$ is the noise error of selecting the $i$-th 2-way marginal, $\phi_i$ is the dependency error of missing $i$, and $x_i\in\{0,1\}$ indicates whether $i$ is selected.

On top of noise error and dependency error, \textsc{DenseMarg} selects the optimal set of 2-way marginals with a greedy algorithm.  
Among the selected 2-way marginals, \textsc{DenseMarg} further merges the overlapping ones whose sizes are small, and derive the final set.

\vspace{2pt} \noindent \textbf{Adding noise to marginals.} 
Gaussian noise is added to each selected marginal to satisfy DP. As proved in ~\cite{zhang2021privsyn} (Theorem 6), a marginal $M$ has a sensitivity of $\Delta_M = 1$, and the Gaussian mechanism dictates that the magnitude of the noise is contingent upon ${\Delta_M}$. 
Consequently, the noisy marginal $M$ is defined as, $\tilde M = M + \mathcal{N}(0, 1/2\rho\mathbf{I})$,
where $\mathcal{N}(0, 1/2\rho\mathbf{I})$ denotes a multi-dimensional random variable following a normal distribution with a mean of 0 and a variance of $1/2\rho\mathbf{I}$.


\vspace{2pt} \noindent \textbf{Marginal post-processing.} 
After the noisy marginals are published, every operation is post-processing, and no extra privacy budget will be consumed~\cite{dp}. \system conducts 3 steps to edit the published marginals to improve their utility.
First, 
we project the invalid distribution into a valid one (i.e., no negative probabilities and their sum equals 1). 
 
Second, when an attribute $f$ is contained by multiple published marginals (e.g., $M_{f,a}$ and $M_{f,b}$), we use weighted average method~\cite{qardaji2014priview} on the marginals to minimize the variance on $f$. 
 
Third, we edit some entries in the marginal tables to make them consistent with protocol rules, e.g., \texttt{byt} has to be larger or equal to \texttt{pkt} (a packet should have at least one byte) and most of FTP packets should use TCP\footnote{Though FTP is supposed to be on TCP, we found exceptions in our data: e.g., on the UGR16 dataset, there are 224 and 1,293 FTP packets (\texttt{dstport} is 21 or 22) using UDP. This could be caused by data collection errors or the abnormal behaviors of clients. So instead of removing the entries that violate protocol rules, we assign a probability threshold $\tau$.}.

\vspace{2pt} \noindent \textbf{Privacy budget allocation.}
Given a privacy budget \(\rho\), which is converted from $(\varepsilon, \delta)$ under Zero Concentrated DP~\cite{zCDP_compose}, we allocate \(0.1\rho\), \(0.1\rho\), and \(0.8\rho\) for data-dependent binning, marginal selection, and publishing noisy marginals.


\subsection{Record Synthesis}
\label{subsec:synthesis}



After the noisy 1-way, 2-way, and combined marginals are derived, this step generates a synthesized dataset that has the same or similar marginals. 
We improve the Gradually Update Method (\textsc{GUM}) developed by PrivSyn for better efficiency, as this step takes most of execution time. 

Specifically, we call our method \textit{GUMMI (GUM with Marginal Initialization)}, which initializes a dataset $D_s$ that contains marginals key to the downstream tasks, e.g., marginals that contain the \textit{label} field as it is essential for flow/packet classification. As such, the correlations between the features (e.g., \texttt{dstport}) and the label are better preserved. We let the data owner specify the key attribute, and let GUMMI select the $n^I$ noisy multi-way marginals that contain attribute,  and orders them by \textit{Pearson standard correlation coefficient} (high to low).
Since the coefficient is computed on the noisy marginals, no privacy budget is consumed at this step. Next, GUM will be applied to iteratively update $D_s$ to replace attributes or duplicate rows, so the the marginals from $D_s$ are close enough to the released marginals.

Finally, each record needs to be decoded due to binning performed in the pre-processing stage. For the most binned fields, we uniformly sample a value within the bin. We also consider the network-related constraints to avoid sampling invalid values (e.g., port number should be less than 65536).

Regarding the timestamp field \texttt{ts}, we leverage the auxiliary field \texttt{tsdiff} generated during pre-processing (see Section~\ref{subsec:pre-processing}) to synthesize its values. Specifically, we first cluster the encoded rows by their identifier, such as IP 5-tuple. Then, we sample within the \texttt{tsdiff} bin range under a Gaussian distribution (rounded to the nearest integer) and add the sampled value to the bin starts.

\section{Evaluation}
\label{sec:evaluation}

In this section,  we first describe the experimental setup, then compare the effectiveness and efficiency of \system with other baseline methods in downstream tasks like packet/flow classification and data sketching. We report the results of attribute-wise measurement, ablation study, and privacy analysis in Appendix~\ref{app:attribute}, ~\ref{app:ablation},  ~\ref{app:mia}.

\subsection{Experimental Setup}
\label{subsec:setup}

\noindent \textbf{Datasets and baseline models.} We select 5 public datasets (UGR16, CIDDS, TON, CAIDA and DC), which are also used by our main baseline model NetShare~\cite{yin2022practical}\footnote{We did not include the Cyber Attack (CA) dataset~\cite{ca} like NetShare, as the original data does not have a label and we did not find another dataset that includes the label attribute.}. These traces are diverse in the deployments, collection logic, and timescales. For baseline models, we consider NetShare, PGM~\cite{PGM} and PrivMRF~\cite{privmrf}. We describe them in Appendix~\ref{app:datasets}. 

\vspace{2pt} \noindent \textbf{Parameters of \system.}
For most experiments, we set the privacy budget $\varepsilon$ equal to 2.0 for \system and the other baseline models, which is a common value used by other works (PrivSyn uses $\varepsilon \in [0.2, 2.0]$ ~\cite{zhang2021privsyn}), providing moderate privacy guarantee. NetShare uses larger $\varepsilon$, from $24.24$ to $10^8$, and we argue that the privacy protection is significantly weakened. In Appendix~\ref{app:ablation}, we test other $\varepsilon$ values.

For \system, we set the maximum number of update iterations during record synthesis to 200. During marginal post-processing, we use $\tau$ to bound the probabilities of certain protocol combinations and we set its value to 0.1.

NetShare requires $\delta$ to be manually configured, we use the same value $10^{-5}$. We also follow its ``DP Pretrained-SAME'' mode, which uses part of its data to pre-train a model and fine-tune the model with the remaining data. 
For PGM, we manually select all 2-way marginals that contain the label attribute of each dataset, which is expected to boost the accuracy on machine-learning based tasks.

\vspace{2pt} \noindent \textbf{Implementations and testing environment.}
We implement \system in Python 3.11.4. 
All the experiments were run on a workstation with 20.04.1-Ubuntu, AMD 3970x CPU (32 cores) and 256GB memory.

\subsection{Data Sketching}
\label{subsec:sketch}


Many network applications leverage sketching to create a compact, efficient data structure for summarizing and analyzing network traffic in real-time. Like NetShare, we consider 4 common sketching algorithms, including Count-Min Sketch (CMS)~\cite{cormode2005improved}, Count Sketch (CS)~\cite{cormode2005improved}, Universal Monitoring (UM)~\cite{liu2016one}, and NitroSketch (NS)~\cite{liu2019nitrosketch}. The threshold for heavy hitters is set to 0.1\%. 

\begin{figure}[!t]
  \centering
  \includegraphics[width=\linewidth]{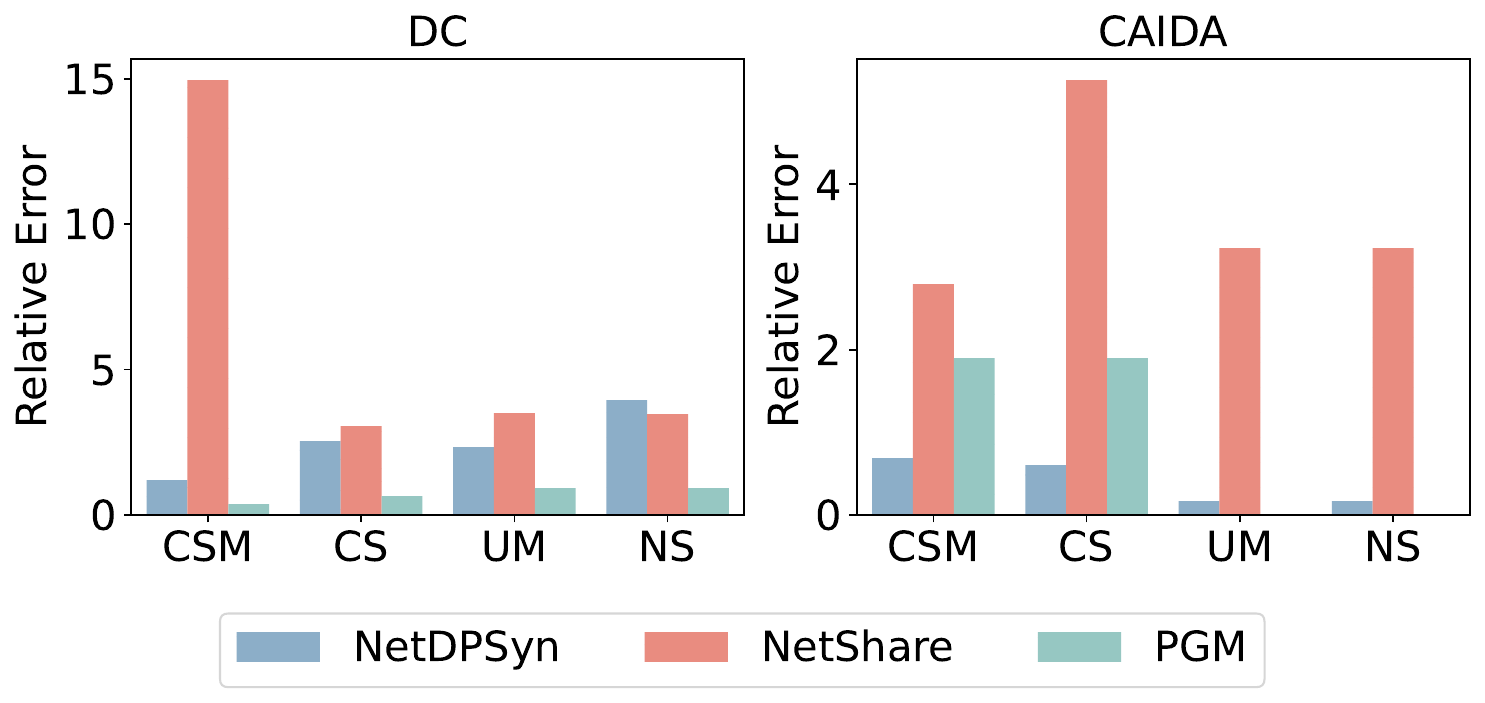}
  \caption{Relative error of various sketch algorithms. The lower the better. }
  \label{fig:sketch}
  \vspace{-10pt}
  
\end{figure}

We compute relative error for heavy hitter count estimation between synthesized and raw data. Assuming the errors for synthesized and raw data are $err_{syn}$ and $err_{raw}$, the relative error is $|\frac{err_{syn}-err_{raw}}{err_{raw}}|$. We use two packet datasets, CAIDA and DC, for this task and compute heavy hitter counts on CAIDA's \texttt{srcip} and DC's \texttt{dstip}. As the sketching algorithms have randomness, we run each sketch 10 times. In Figure~\ref{fig:sketch}, we show the results of 4 sketching methods on the datasets, and NetShare is significantly worse than the other methods, except for NitroSketch on DC. In fact, NetShare performs particularly worse for simpler sketching algorithms, like CSM on DC (12x relative error compared to \system) and CS on CAIDA (9x relative error compared to \system).  \system outperforms PGM in CAIDA when CSM and CS are used. 
For PrivMRF, we found it cannot run on DC and CAIDA as it exceeds our machine's memory.

\subsection{Machine-learning Tasks}
\label{subsec:ml}


Next, we consider machine-learning tasks, which highly depend on feature engineering, to assess whether the attribute correlations are preserved. 

\vspace{2pt} \noindent \textbf{Classification on flows.}
Anomaly detection is a key use case on network flows. We consider 3 datasets, including TON, UGR16, and CIDDS, for this task. For TON, we use its \texttt{type} attribute as the classification label, which describes the attack type (10 categorical values including ``normal'', ``ddos'', etc.). For UGR16 and CIDDS, we use their binary \texttt{label} attribute (benign or malicious) as the classification label. 
All attributes except the classification label are used as features. 
We randomly split each data into 80\% for training and 20\% for testing\footnote{NetShare claimed to split the data by time orders in the paper. We initially followed the same data split method for TON, but the classification accuracy even on raw data is low. We found most of the simulated attacks in TON happen at the end of the period, so the training data will not contain sufficient attack records if splitting by time orders.} and we measured classification accuracy, which is defined as $\frac{TP + TN}{TP + TN + FP + FN}$, where TP, TN, FP and FN are true positives, true negatives, false positives and false negatives.
We implemented five common models: Decision Tree (DT), Logistic Regression (LR), Random Forest (RF), Gradient Boosting (GB), and Multi-layer Perceptron (MLP). 

\begin{figure*}[ht]
  \centering
  \includegraphics[width=\textwidth]{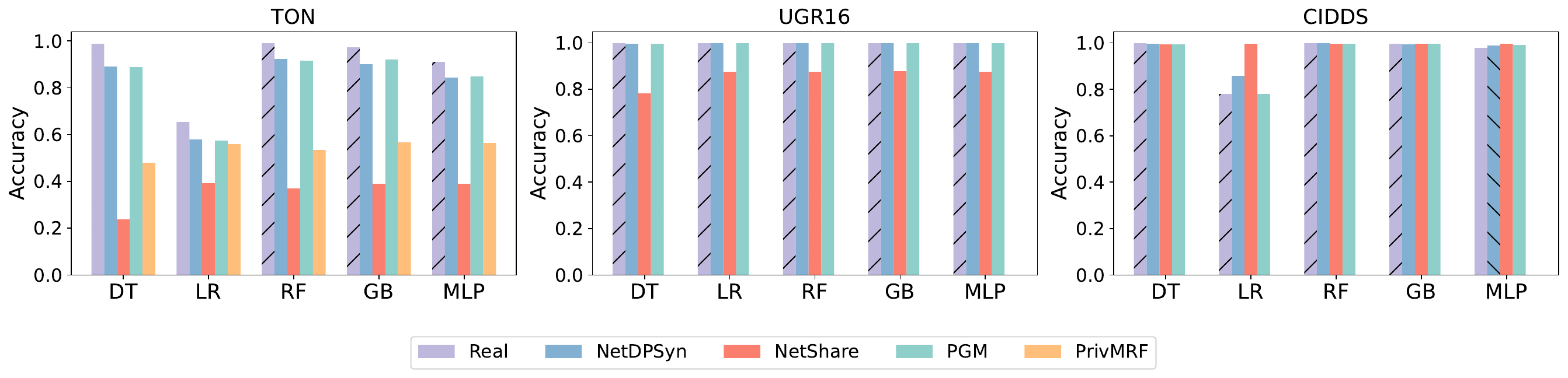}
  \caption{Classification accuracy of three flow datasets. The higher the better.}
  \label{fig:classification}
  \vspace{-10pt}
\end{figure*}

Figure~\ref{fig:classification} compares the classification accuracy of the 3 datasets. The performance gap on TON is most prominent, as the data generated by PGM and \system lead to close accuracy as the raw data (e.g., 0.987 for Raw data, 0.889 for \system, 0.886 for PGM with DT), but the data from NetShare lead to significantly lower accuracy (e.g., 0.235 with DT). The accuracy with LR is low for all models, mainly due to the simplicity of LR. 
For UGR16 and CIDDS, the data generated by most methods lead to very high accuracy, which is close to the accuracy on raw data. This is mainly because the classification label is binary and the data is highly imbalanced,
so achieving high accuracy is much easier (e.g., even predicting every row as benign on UGR16 achieves 0.997 accuracy).
Still, the data from NetShare lead to 0.1 to 0.2 lower accuracy on UGR16.

Instead of always achieving high accuracy, it is more important that a classification model achieves similar accuracy on raw and synthesized datasets for fidelity, even if the accuracy is low in both cases. As such, we compare the rankings of the 5 models when they are trained/tested on raw/synthetic data. Like NetShare, we compute Spearman’s rank correlation coefficient between the raw/synthetic data. Table~\ref{tbl:classification} shows that \system achieves the highest coefficient.

We also found \system achieves similar or even better data utility than NetShare \textit{without DP}. For example, the rank correlations are 0.70 TON and 0.90 CIDDS for NetShare (Table 3~\cite{yin2022practical}), while 0.90 and 0.90 for \system (Table~\ref{tbl:classification}).



\begin{table}[!t]
\centering
\caption{Spearman’s rank correlation of prediction algorithms on TON, CIDDS and UGR16. The higher the better. ``N/A'' for PrivMRF at CIDDS and UGR16 as memory is exceeded.}
\begin{tabular}{lcccc}
\hline
& \system & NetShare & PGM & PrivMRF \\
\hline
TON   & 0.90 & -0.90 & 0.70 & -0.50 \\
CIDDS & 0.90 & -0.30 & 0.70 & N/A \\
UGR16 & 0.45 & 0.36 & 0.36 & N/A \\
\hline
  \vspace{-10pt}
\end{tabular}
\label{tbl:classification}
\end{table}

\vspace{2pt} \noindent \textbf{Anomaly detection on packets.}
For packet datasets, we leveraged NetML~\cite{yang2020comparative}, an open-source library to generate flow representations and leverage  its the default one-class support vector machine (OCSVM) to detect abnormal packets. Like NetShare, we choose 5 modes of NetML: IAT, SIZE, IAT\_SIZE, STATS, SAMP-NUMP (SN) and SAMP-SIZE (SS). As an example, STATS contains 10 statistical features from a flow, like flow duration, number of packets per second, etc. 

\begin{figure}[ht]
  \centering
  \includegraphics[width=\linewidth]{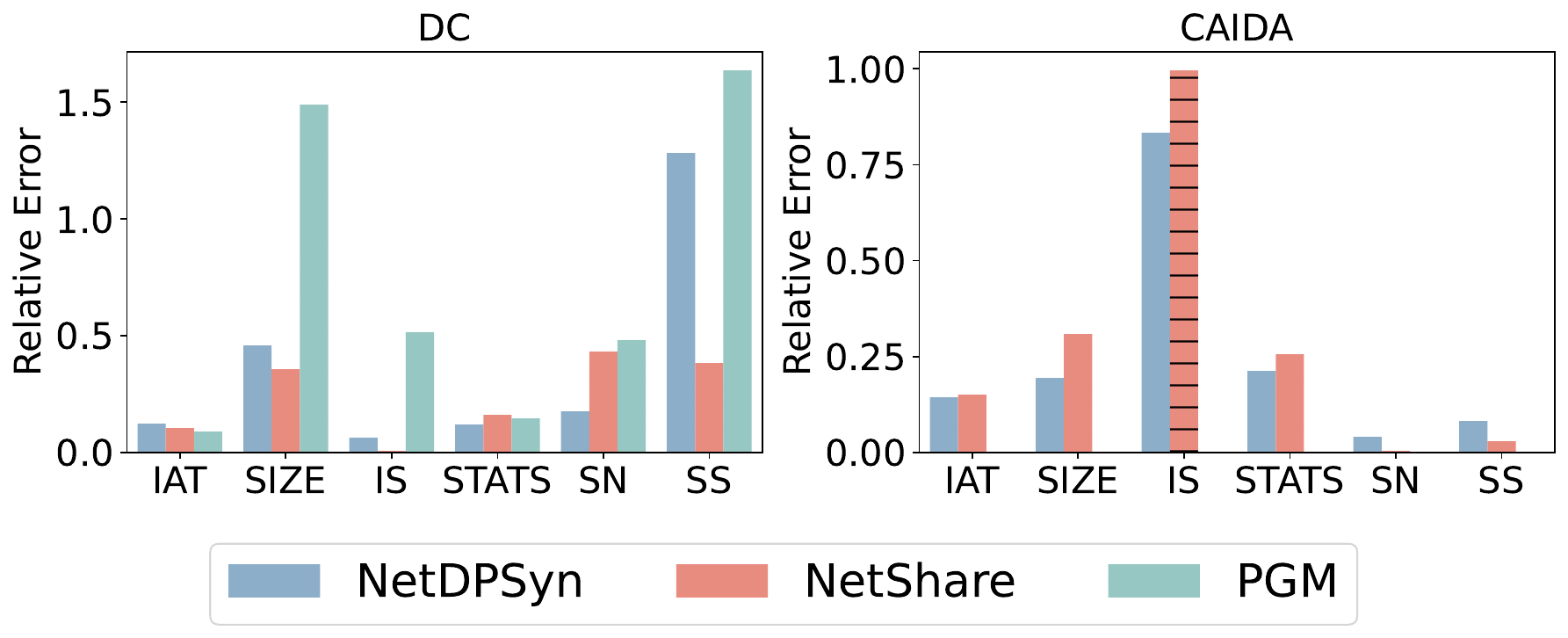}
  \caption{NetML results on two packet datasets. The lower the better.
  } 
   
  \label{fig:netml}
  \vspace{-10pt}
\end{figure}

For each round of running OCSVM, we obtain an anomaly ratio for the raw data ($ano_{raw}$) and one for the synthesized data ($ano_{syn}$). We compute the relative error as $|\frac{ano_{syn}-ano_{raw}}{ano_{raw}}|$. As NetML only accepts flows with at least two packets, a subset of packets can be used. Figure~\ref{fig:netml} shows the results. 
In addition to PrivMRF being unable to run on the two packet datasets, we found that PGM also encounters ``NaN'' on CAIDA. 
This is because
in the dataset generated by PGM, only a few number of flows contain two packets. 
We found \system has comparable results as NetShare except SS. PGM has a very high relative error for SIZE and SS.

We also compute the Spearman’s rank correlation coefficient between the raw/synthetic data, and the result is shown in Table~\ref{tbl:netml}. \system performs best.


\begin{table}[!t]
\centering
\caption{Rank correlation of NetML model for packet anomaly detection. The higher the better.}
\begin{tabular}{lcccc}
\hline
& \system & NetShare & PGM & PrivMRF \\
\hline
CAIDA   & -0.48  &  -0.82 &  N/A  & N/A \\
DC &  0.26 &  -0.65 &-0.26 & N/A \\
 
\hline
\end{tabular}
\vspace{-5pt}
\label{tbl:netml}
\end{table}

\subsection{Efficiency}
\label{subsec:ablation}

We compare the efficiency of different methods by measuring their running time, as shown in Table \ref{tbl:efficiency}. \system is much faster than NetShare and PGM in most cases, e.g., it is 5x and 2x faster than Netshare and PGM in processing CIDDS. PrivMRF is particularly slow, e.g., taking 240 minutes to synthesize TON, and it even exceeds the memory limit when processing datasets larger than TON. This is because PrivMRF selects too many marginals.

\begin{table}[!t]
\centering
\caption{Running Time of each method in minutes.}
\begin{tabular}{lcccc}
\hline
& \system & NetShare & PGM & PrivMRF \\
\hline
TON   & 10 & 27 & 70 & 240\\
CIDDS & 20 & 100 & 55 & N/A \\
UGR16 & 40 & 94 & 55 & N/A \\
CAIDA & 35 &  30 & 54 & N/A \\
DC & 20 & 100 & 24 & N/A \\
\hline
\end{tabular}
\vspace{-5pt}
\label{tbl:efficiency}
\end{table}

 \section{Conclusion and Future Works}

This paper introduces \system, a method for synthesizing high-fidelity network traces with Differential Privacy. We take a drastically different path from prior works based on generative models like NetShare. Specifically, \system uses DP as the core to capture the underlying distribution of the high-dimensional network traces and synthesize records from them under DP's post-processing property. Our initial results are promising, showing significantly better data fidelity and efficiency than baseline methods like NetShare, PGM and PrivMRF, under the same DP privacy budget.

\vspace{2pt} \noindent \textbf{Limitations and future works.} 
We consider our work as the first step, and we acknowledge a few limitations as listed below. 1) We model the temporal patterns with packet-arrival intervals, it is relatively coarse representation. NetShare might outperform \system in downstream tasks that require complex temporal modeling. Combining \system with recurrent neural networks (RNN) might lead to better results in this aspect. 2) \system does not cover all types of network environment and data types (e.g., payload data). 3) \system performs well when the number of attributes is relatively small (e.g. all tested datasets have no more than 15 attributes). High-dimensional data might lead to computational and memory inefficiencies, and we can integrate dimensionality reduction techniques to address this issue. 4) We did not evaluate advanced downstream tasks like graph-based anomaly detection~\cite{oprea2015detection}, which could be an interesting problem to study in the future.


    \section*{Acknowledgement}
We thank our reviewers and shepherd for the valuable suggestions. We thank Rui Zhao for his contribution in the initial system implementation. This project is supported under NSF CNS-2220434 and CNS-2220433. 



 \bibliographystyle{ACM-Reference-Format}
 \balance
 \bibliography{refs}

\appendix

\section{Ethics}

We evaluated \system with the common downstream tasks like machine-learning-based classification and data sketching, using the same set of public datasets as NetShare. We do not foresee any ethical issue in this aspect. A potential concern is about the inference attack we conducted to evaluate the privacy leakage of the raw and synthesized data. We follow a similar attack procedure as Chanyaswad et al.~\cite{chanyaswad2019ron}, focusing on assessing system vulnerabilities rather than attempting to de-anonymize any data. Only aggregated statistics like attack accuracy is derived. It is important to note that our intent was to underline the importance of robust privacy protections, not to compromise any individual’s anonymity.

\section{Workflow of \system}
\label{app:workflow}

In Algorithm~\ref{alg:system}, we show workflow of \system in pseudo-code, consisting of pre-processing, marginal selection, noise addition, and record synthesis.

\begin{algorithm}[!h]
\caption{Pseudo-code of \system.}
\begin{algorithmic}[1]
\Require Private dataset $Do$, privacy budget $\rho$
\Ensure Synthetic dataset $Ds$
\State Binning each attribute using type-dependent binning method
\State Add the auxiliary temporal attribute
\State Publish 1-way marginals using GM with $\rho_1 = 0.1\rho$
\State Binning each attribute with frequency-dependant method
\State Select 2-way marginals with $\rho_2 = 0.1\rho$
\State Combine marginals with small sizes
\State Publish combined marginals using Gaussian Mechanism with $\rho_3 = 0.8\rho$
\State Make noisy marginals consistent on the sum of cell values, shared attributes, and protocol rules 
\State Construct encoded dataset $D_e$ using GUMMI from an initialized dataset $D_s$
\State Decode $D_e$ by value sampling within bins
\State Reconstruct the timestamp attribute with the auxiliary temporal attribute
\end{algorithmic}
\label{alg:system}
\end{algorithm}

\section{Examples of Marginal Tables}
\label{app:marginals}

In Table~\ref{tbl:marginals}, we show examples of 1-way and 2-way marginal tables and their noisy versions.

\begin{table}[!h]
\centering

\begin{subtable}[t]{.5\linewidth}
\centering
\begin{tabular}{cc}
\toprule
$v$ & $M_{\text{d}}(v)$ \\
\midrule
$\langle\text{53, *}\rangle$ & 82828 \\
$\langle\text{80, *}\rangle$ & 68748 \\
$\langle\text{15600, *}\rangle$ & 27255 \\
\bottomrule
\end{tabular}
\caption{1-way marginal for \texttt{dstport}.}
\end{subtable}%
\begin{subtable}[t]{.5\linewidth}
\centering
\begin{tabular}{cc}
\toprule
$v$ & $M_{\text{t}}(v)$ \\
\midrule
$\langle*, \text{normal}\rangle$ & 166494 \\
$\langle*, \text{injection}\rangle$ & 15951 \\
\bottomrule
\end{tabular}
\caption{1-way marginal for \texttt{type}.}
\end{subtable}

\bigskip 

\begin{subtable}[t]{.5\linewidth}
\centering
\begin{tabular}{cc}
\toprule
$v$ & $M_{\text{dt}}(v)$ \\
\midrule
$\langle\text{53, normal}\rangle$ & 74547.08\\
$\langle\text{53, injection}\rangle$ & 554.71 \\
$\langle\text{80, normal}\rangle$ & 12297.88 \\
$\langle\text{80, injection}\rangle$ & 15396.66 \\
$\langle\text{15600, normal}\rangle$ & 27247.02 \\
$\langle\text{15600, injection}\rangle$ & 20.09 \\
\bottomrule
\end{tabular}
\caption{2-way noisy marginal before marginal post-processing.}
\end{subtable}%
\begin{subtable}[t]{0.56\linewidth}
\centering
\begin{tabular}{cc}
\toprule
$v$ & $\tilde{M}_{\text{dt}}(v)$ \\
\midrule
$\langle\text{53, normal}\rangle$ & 74566 \\
$\langle\text{53, injection}\rangle$ & 558 \\
$\langle\text{80, normal}\rangle$ & 12308\\
$\langle\text{80, injection}\rangle$ & 15364 \\
$\langle\text{15600, normal}\rangle$ & 27255 \\
$\langle\text{15600, injection}\rangle$ & 0 \\
\bottomrule
\end{tabular}
\caption{Actual 2-way marginal.}
\end{subtable}
\caption{Marginal tables for \texttt{dstport} and \texttt{type} computed on TON dataset. Due to space limit, only the first few rows are shown.}
\label{tbl:marginals}
\end{table}

\section{Datasets and Baseline Models}
\label{app:datasets}




We obtained copies of datasets from the authors of NetShare, which are subsets of original datasets.
In Table~\ref{tbl:datasets}, we describe the basic statistics of each dataset.
\begin{itemize}[leftmargin=*]
    \item \textbf{URG16}~\cite{UGR}: This dataset comprises network traffic collected from NetFlow v9 collectors within a Spanish ISP's network, including various attacks. The specific data was gathered during the third week of March 2016.
    \item \textbf{CIDDS}~\cite{ring2017creation}: The dataset replicates a small business environment featuring various clients and servers (such as email and web services) into which malicious traffic has been intentionally introduced. Each NetFlow entry is meticulously recorded, classified as benign or an attack, and categorized by attack type, including DoS, brute force, and port scans. 
    \item \textbf{TON\_IoT (TON)}~\cite{moustafa2021new}: TON is a collection representing telemetry from IoT sensors. Our evaluations focus on a subset of this dataset named ``Train\_Test\_datasets.'' Cyber-attacks, such as backdoor, DDoS, and injection, are simulated.
    \item \textbf{CAIDA}~\cite{caida}: This dataset has anonymized data gathered from high-speed monitors located on a commercial backbone network.
    \item \textbf{Data Center (DC)}~\cite{dc}: It is a collection of packet captures from the "UNI1" data center, which is used by~\cite{benson2010network}.
\end{itemize}


\begin{table}[!h]
\centering
\caption{Summary of datasets used in our experiments. Domain is computed by summing the domain sizes from all attributes.}
\resizebox{0.48\textwidth}{!}{
\begin{tabular}{l|c|c|c|c|l}
\hline
\textbf{Dataset} & \textbf{Records} & \textbf{Attributes} & \textbf{Domain} & 
\textbf{Label} &
\textbf{Type} \\
\hline
TON & 295,497 & 11 & $2 \cdot 10^{6}$ & type & flow \\
UGR16 & 1,000,000 & 10 & $4 \cdot 10^{6}$ & type & flow\\
CIDDS & 1,000,000 & 11 & $6 \cdot 10^{6}$ & type & flow\\
CAIDA & 1,000,000 & 15 & $1 \cdot 10^{7}$ & flag & packet\\
DC & 1,000,000 & 15 & $1 \cdot 10^{7}$ & flag & packet\\
\hline
\end{tabular}
}
\label{tbl:datasets}
\end{table}

Regarding baseline models to compare with \system, we mainly use the GAN-based NetShare. 
We also consider PGM~\cite{PGM} and PrivMRF~\cite{privmrf}, which are two other marginal-based synthesis approaches. These methods are briefly described below.
\begin{itemize}[noitemsep,topsep=0pt]
    \item  \textbf{GAN-based  NetShare~\cite{yin2022practical}.} It uses a time-series generator to generate packets' metadata and their measurements, and then uses one discriminator to differentiate the packet time series and another auxiliary discriminator to discriminate only on metadata. To better capture the header field correlation spanning multiple packets or epochs, NetShare splits the network data by flows and uses the time-series GAN to synthesize new flows in parallel. To fairly compare with \system and other baseline models, we use its DP version that applies DP-SGD when training the time-series GAN. 
    We use its code from~\cite{netshare-code}.

    \item  \textbf{PGM~\cite{PGM}.} This approach concurrently selects marginal distributions and establishes the Bayesian network's structure. It does so by iteratively optimizing the information gain using the exponential mechanism. Following this process, synthetic data is generated by sampling from the joint distribution derived from the established topology of the Bayesian network. 
    The major limitation of PGM is that the operator needs to manually provide a list of marginals for PGM to synthesize from. 
    We use its code from~\cite{pgm-code}.

    \item \textbf{PrivMRF~\cite{privmrf}.} This method addresses the limitations of PGM by using a DP algorithm to automatically select a set of low-dimensional marginals. Selected marginals are used to construct a Markov random field (MRF), which models dependencies among attributes in the input data. The MRF then serves as the basis for generating synthetic data. 
    We use its code from~\cite{privmrf-code}.
\end{itemize}

\begin{figure}[h]
  \centering
  \includegraphics[width=\linewidth]{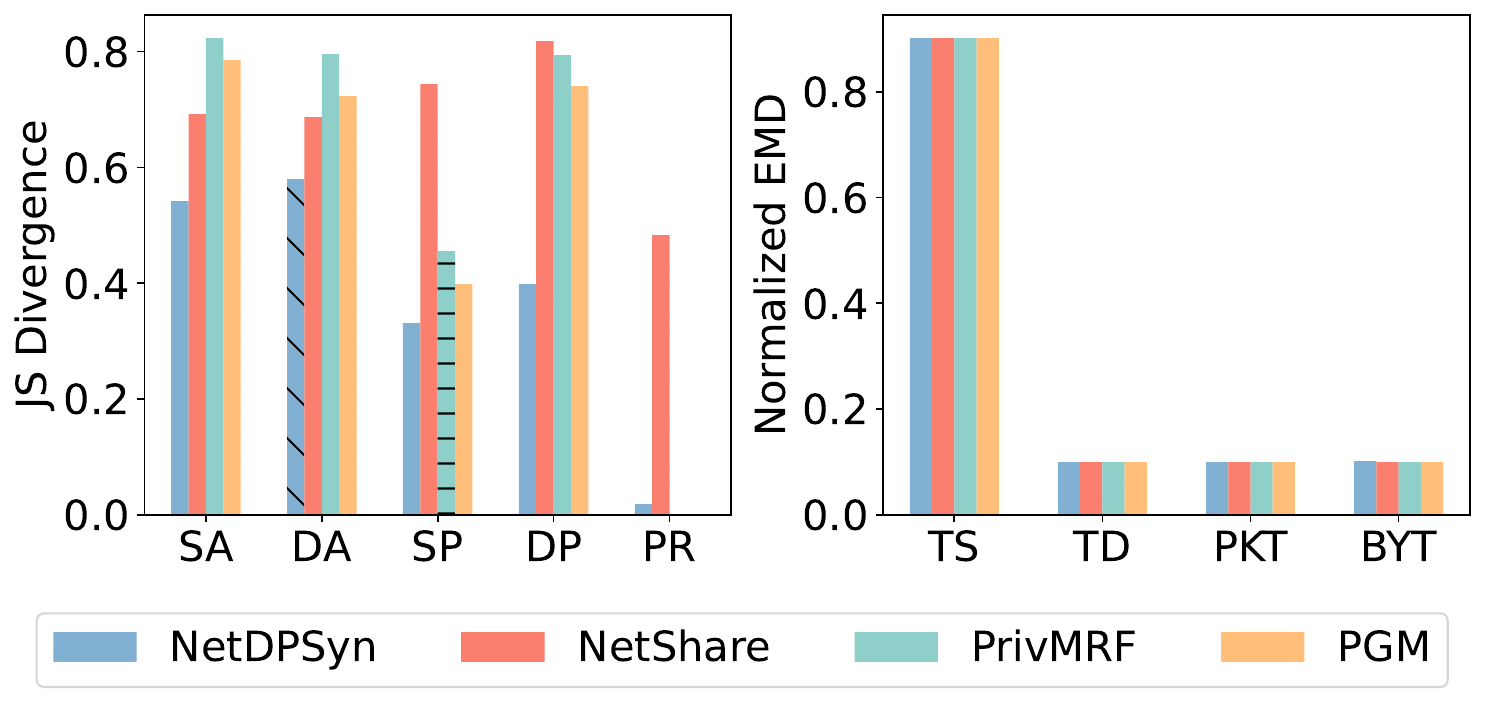}
  \caption{TON (NetFlow) JSD and EMD. The lower the better.
  }
  \label{fig:ton_attr}
  \vspace{-10pt}
\end{figure}

\begin{figure}[h]
  \centering
  \includegraphics[width=\linewidth]{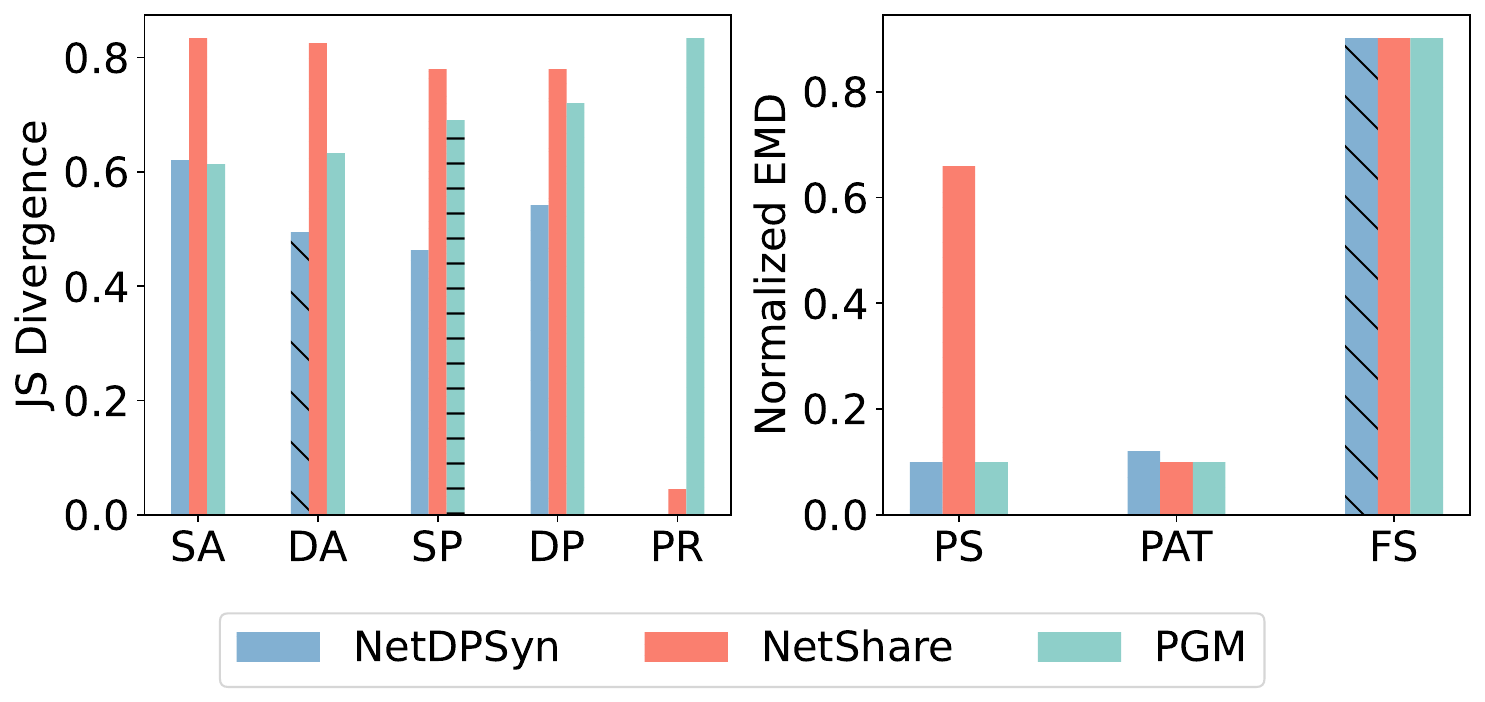}
  \caption{CAIDA (Packet) JSD and EMD. The lower the better.}
  \label{fig:caida_attr}
  \vspace{-10pt}
  
\end{figure}

\section{Attribute-wise Measurement}
\label{app:attribute}

We measure the value distribution of individual attributes after synthesis. Like NetShare, for categorical attributes, we use Jensen-Shannon divergence (JSD) to measure the distance between the synthesized attribute and the raw attribute. For continuous attributes, we use Earth Mover’s Distance (EMD). Because different attributes have vastly different EMD ranges, we normalize the EMDs to $[0.1,0.9]$ for better figure readability.

Regarding categorical attributes, for both network flows and packets, we compute 5 metrics, including SA (relative frequency of \texttt{srcip} ranking in a descending way), DA (same metric for \texttt{dstip}), SP (port number distribution of \texttt{srcport}, ranging from 0 to 65535), DP (same metric for \texttt{dstport}), and PR (relative frequency of \texttt{proto}). 

Regarding continuous attributes, the metrics are different for network flows and packets. For flows, we directly compute EMD on \texttt{ts}, \texttt{td}, \texttt{pkt}, and \texttt{byt}. \texttt{ts} and \texttt{td} are in milliseconds.
For packets, we compute PS (Packet Size in bytes, same as \texttt{pkt\_len}), PAT (Packet Arrival Time in milliseconds, same as \texttt{ts}), and FS (Flow Size, or the number of packets under an IP 5-tuple). 
The explanations of the attributes are in Section~\ref{subsec:network}.

In Figure~\ref{fig:ton_attr} and Figure~\ref{fig:caida_attr}, we show the results on one network flow dataset (TON) and one packet dataset (CAIDA). For the categorical metrics in TON, \system is consistently better than the other methods, with 30\%-45\% lower JSD.
NetShare performs notably worse in PR, when the other methods have close to zero JSD.
Achieving low JSD for PR should be relatively easy, as it only has 3 categorical values, TCP, UDP, and ICMP, but the noises amplified under DP-SGD (explained in Section~\ref{subsec:motivation}) significantly degrade the data fidelity after synthesized with NetShare.
The normalized EMDs are all similar, due to the raw EMDs are either very large (so close to 0.9) or very small (so close to 0.10). For CAIDA, \system is only slightly worse than NetShare for PAT, and we speculate the time-series GAN used by NetShare offsets the noises added by DP-SGD. Notice that PrivMRF is not shown in Figure~\ref{fig:caida_attr}, as it exceeds our memory limit.

\begin{figure}[h]
  \centering
  \includegraphics[width=\linewidth]{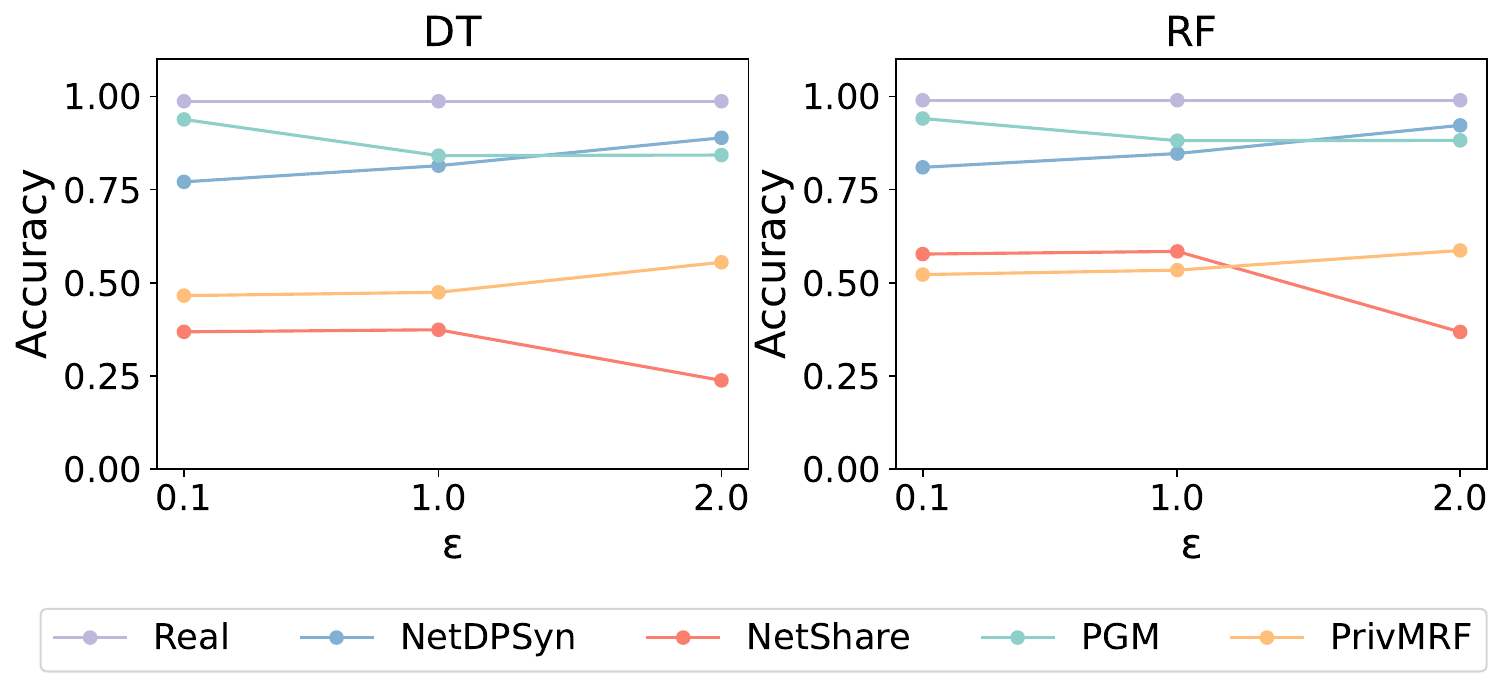}
  \caption{TON (NetFlow) accuracy with different $\varepsilon$ value. The higher the better.}
  \label{fig:ton_attr_0.1}
\end{figure}

\section{Ablation Study}
\label{app:ablation}

\vspace{2pt} \noindent \textbf{Noise scale.} 
In the default setting, we choose $\varepsilon=2.0$ and run all the experiments. Here, we reduce $\varepsilon$ to 1.0 and 0.1, which means the privacy protection becomes stronger, to assess its impact on the data fidelity. In 
Figure~\ref{fig:ton_attr_0.1} shows the accuracy under different $\varepsilon$ on DT and RF measurements of TON. The result shows that \system achieves consistent accuracy with much lower $\varepsilon$ with DT and RF.




Next, we conduct a more comprehensive comparison between \system and NetShare, with a much larger range of $\varepsilon$ (4.0, 8.0, 16.0, 32.0, 64.0, $10^3$ and $10^{10}$).
We are interested in whether NetShare is able to match the performance of \system with very large $\varepsilon$.
In Table~\ref{table:ton_diff_eps}, we show the RF and DT classification accuracy on TON. For \system, its accuracy climbs up to 0.94 after $\varepsilon$ reaches to 16.0. For NetShare, when $\varepsilon = 10^{10}$, accuracy of DT and RF accuracy increase noticeably, but they do not exceed 0.4 and 0.6 under very large $\varepsilon$ ($10^{10}$). Noticeably, NetShare without DP has around 0.6 accuracy on TON dataset (Figure 12 in ~\cite{yin2022practical}). 
 In Table~\ref{table:ugr16_diff_eps}, we show the result on another dataset UGR16, but the accuracy does not change much under very large  $\varepsilon$.


\begin{table}[h]
\centering
\begin{tabular}{cc|c|c|c}
\toprule
& \multicolumn{2}{c}{\textbf{DT}} &  \multicolumn{2}{c}{\textbf{RF}}\\
\midrule
$\varepsilon$ & \system & NetShare & \system & NetShare\\
\midrule
4.0 & 0.910 & 0.213 & 0.932 & 0.368\\
16.0 & 0.941 & 0.235 & 0.954 & 0.389\\
32.0 & 0.943 & 0.257 & 0.955 & 0.413\\
64.0 & 0.946 & 0.258 & 0.955 & 0.427\\
$10^{3}$ & 0.947 & 0.260 & 0.956 & 0.423\\
$10^{10}$ & 0.948 & 0.389 & 0.957  & 0.580\\
\bottomrule
\end{tabular}
\caption{Comparison of TON (NetFlow) accuracy between \system and NetShare with a large range of $\varepsilon$.}
\label{table:ton_diff_eps}
\end{table}

\begin{table}[h]
\centering
\begin{tabular}{cc|c|c|c}
\toprule
& \multicolumn{2}{c}{\textbf{DT}} &  \multicolumn{2}{c}{\textbf{RF}}\\
\midrule
$\varepsilon$ & \system & NetShare & \system & NetShare\\
\midrule
4.0 & 0.976 & 0.779 & 0.986 & 0.870\\
16.0 & 0.978 & 0.781 & 0.987 & 0.874\\
32.0 & 0.980 & 0.781 & 0.988 & 0.875\\
64.0 & 0.983 & 0.782 & 0.988 & 0.878\\
$10^{3}$ & 0.986  &  0.783  & 0.990  & 0.878 \\
$10^{10}$ & 0.989  & 0.783   & 0.992  & 0.879 \\
\bottomrule
\end{tabular}
\caption{Comparison of UGR16 (NetFlow) accuracy between \system and NetShare with a large range of $\varepsilon$.}
\label{table:ugr16_diff_eps}
\end{table}

\vspace{2pt} \noindent \textbf{Comparison between GUMMI and GUM.} 
For record synthesis, we propose GUMMI to improve the efficiency. The key motivation is that when the update iteration is set to 200 (the default value used by PrivSyn), GUM consumed approximately \textit{90\%} of the total time for a single experiment run. 

\begin{figure}[t]
  \centering
  \includegraphics[width=\linewidth]{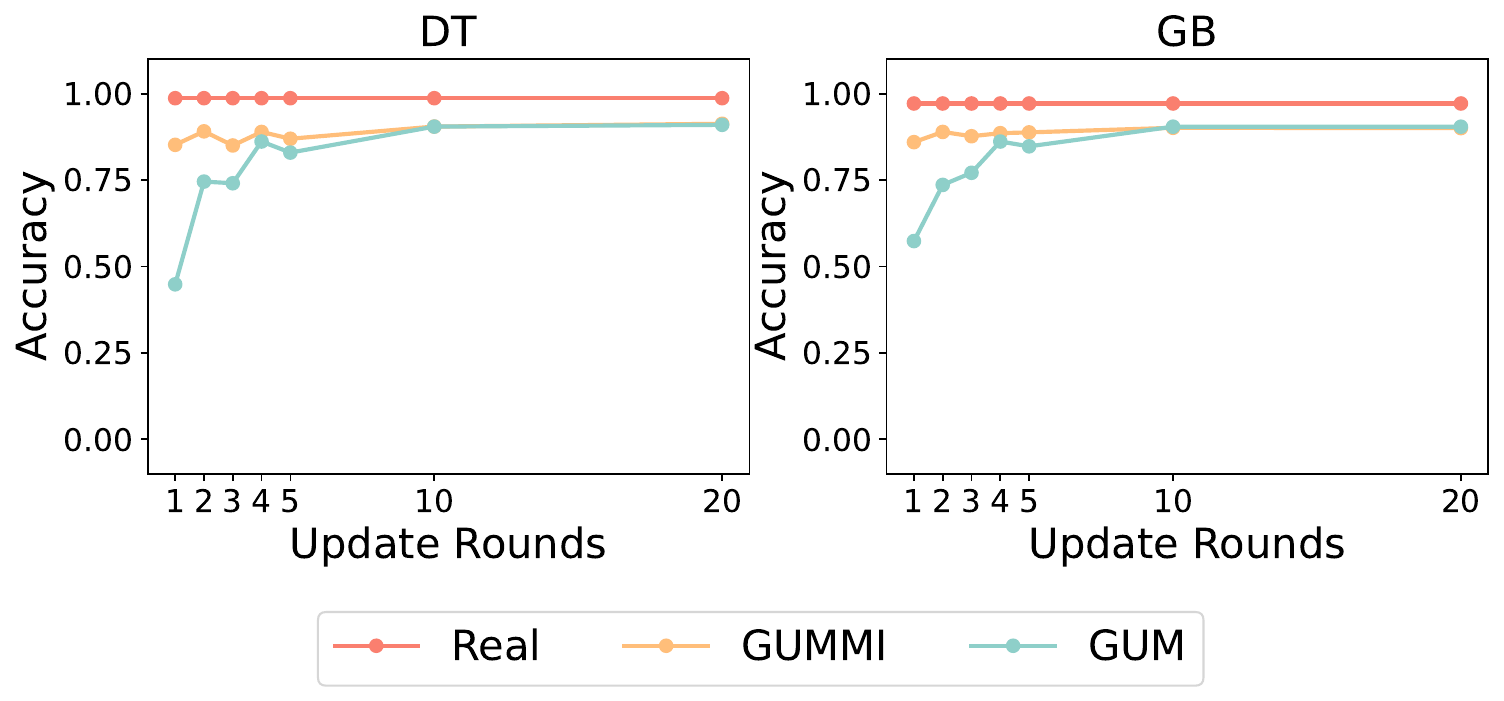}
  \caption{Comparison between GUMMI and GUM in update iterations from 1 to 20. The classification accuracies of DT and GB on TON are shown.} 
  \label{fig:gummi}
\end{figure}

Here, we choose the task of classification on TON as an example to compare GUMMI and GUM. We choose 7 values for update iterations, \{1, 2, 3, 4, 5, 10, 20\}, and run \system with GUMMI and GUM separately. The classification accuracy corresponding to each setup is shown in Figure \ref{fig:gummi}.
At the initial update iterations, the accuracy observed under GUM is significantly lower than GUMMI (0.45 vs 0.85 for the decision tree). The accuracy is similar after 10 update rounds. 
As such, when record synthesis takes too long, especially for large network datasets, the operator can choose smaller update iteration numbers to obtain the results faster. This is especially helpful when the operator needs to search all possible parameters to find the best privacy-utility tradeoff in synthesis.   

\section{Privacy Analysis under Inference Attacks} 
\label{app:mia}

We follow the basic MIA method~\cite{yeom2018privacy} to attack the models (i.e., classifiers) trained on the raw dataset of TON, and we achieve 64.01\% attack accuracy. We also tested MIA on the TON dataset synthesized by \system, and the attack accuracy drops to 55.87\% at $\varepsilon=2$. Under $\varepsilon=0.1$, it further drops to 40.85\%. 
Since \system follows record-level DP, for a synthesized packet dataset, it provides per-packet guarantee, which might not offer practical privacy guarantee. Different DP notions might be needed, e.g., user-level DP~\cite{ghazi2021user}, and we leave the analysis as a future work.

Chanyaswad et al. ~\cite{chanyaswad2019ron} conducted a similar privacy analysis by computing attack accuracy under MIA on the raw and their synthesized dataset (Realistic Sensor Displacement dataset). Similar results were obtained (around 72\% and 50\% attack accuracy on raw and $\varepsilon=2$ synthesized data, as shown in Figure 2 of ~\cite{chanyaswad2019ron}), and they conclude that data synthesis is effective to contain privacy leakage. 
\textcolor{red}{}{Yet, we acknowledge that the defense can be weakened under more powerful attacks in the real-world setting~\cite{meeus2023achilles, carlini2022membership}.
}

\end{document}